\documentclass[aps,pra,superscriptaddress,amsmath,amssymb,preprintnumbers,showpacs]{revtex4}
\usepackage{color}
\usepackage{bm}

\begin{document}
\title{Estimation of the repeatedly-projected reduced density matrix under decoherence}

\author{H. Nakazato}\email{hiromici[at]waseda.jp}
\affiliation{Department of Physics, Waseda University, Tokyo 169-8555, Japan}

\author{K. Yuasa}\email{yuasa[at]hep.phys.waseda.ac.jp}
\affiliation{Waseda Institute for Advanced Study, Tokyo 169-8050, Japan}
\affiliation{Department of Physics, Waseda University, Tokyo 169-8555, Japan}

\author{B. Militello}\email{bdmilite[at]fisica.unipa.it}
\affiliation{MIUR and Dipartimento di Scienze Fisiche ed Astronomiche dell'Universit\`{a} di Palermo, Via Archirafi, 36, I-90123 Palermo, Italy}

\author{A. Messina}
\affiliation{MIUR and Dipartimento di Scienze Fisiche ed Astronomiche dell'Universit\`{a} di Palermo, Via Archirafi, 36, I-90123 Palermo, Italy}

\date[]{March 26, 2008}
\begin{abstract}
Decoherence is believed to deteriorate the ability of a
purification scheme that is based on the idea of driving a system
to a pure state by repeatedly measuring another system in
interaction with the former and hinder for a pure state to be
extracted asymptotically. Nevertheless, we
find a way out of this difficulty by deriving an analytic
expression of the reduced density matrix for a two-qubit
system immersed in a bath. It is shown that we can still extract a
pure state if the environment brings about only dephasing effects.
In addition, for a dissipative environment, there is a
possibility of obtaining a dominant pure state when we perform a
finite number of measurements.
\end{abstract}

\maketitle

\section{Introduction}\label{sec:introduction}
The notion of measurement constitutes a key element in the quantum
theory and there is a long history of debate on this
subject~\cite{ref:qmt}. It is, however, relatively recent that the
following fact has been confirmed in a real laboratory that the
measurement, which is usually described by a projection operator
for simplicity but has to be replaced, in a rigorous sense, with a
generalized spectrum decomposition~\cite{ref:GSD}, does affect the
dynamics of the quantum system drastically and in an essential
way. The quantum Zeno effect~\cite{ref:QZEreview}, the effect
caused by frequently repeated measurements and resulting in
hindrance of the dynamical evolution of the quantum
system~\cite{ref:qd} just measured, is one of such typical and
well-known examples and has been discussed quite intensively after
its first experimental observation~\cite{ref:QZE}. It has also
become clear that the action of measurement can have much more
profound effects on the quantum systems than one naively expects
or imagines from such a phrase like ``wave function collapse.''
Indeed, the effects have been shown to transfer to other quantum
systems, not directly measured, but in interaction with the system
measured, and the action of measurement can be utilized to drive
even such quantum systems not directly touched to a pure state
irrespective of their initial mixed states. This is the essence of
the proposal of the repeated-measurement-based purification
scheme~\cite{ref:Nakazato_PRL}, the applicability and
effectiveness of which have been examined in various
cases~\cite{ref:p_appl}.

When one considers a possible implementation of such a theoretical
scheme to experiments in the laboratories, it is necessary to
examine its robustness against various imperfections to be
encountered in the experiments, which may be expressed as a sort
of decoherence from the environment. According to the analysis in
Ref.~\cite{ref:PrevPaper}, the projected dynamics of the quantum
system suffers from the decoherence effect and it is shown that
such an effect can deteriorate the ability of the purification
scheme to such an extent that no (nontrivial) pure state is able
to be extracted in the limit of infinite number of projections.
While this is somewhat an anticipated and discouraging result, the
numerical estimations of the purity of the state to be extracted
for a qubit system, which is coupled with another qubit that is
subject to repeated interrogations and is immersed together with
the latter in a common bosonic bath, show that there are parameter
regions where states with high purities can be
extracted~\cite{ref:PrevPaper}. On this basis we have studied the
possibility to extract a quantum state with a higher purity under
a dissipative environment by this purification scheme, even though
it can not survive in the limit of infinite number of measurements
because of a kind of no-go theorem~\cite{ref:PrevPaper}. This
paper is devoted to explore further such a possibility of
extracting a dominant pure state at a finite number of
measurements under a dissipative environment, as well as to point
out its robustness against a dephasing environment, at infinite
number of measurements.

It is desirable to obtain an explicit expression of the reduced
density matrix for such investigations, though in the previous
study~\cite{ref:PrevPaper} the ability of the purification scheme
is discussed exclusively in terms of purity. Since solving
analytically the projected dynamics with decoherence, that is the
evolution described by a master equation supplemented with
projections, is in general quite difficult, we confine ourselves
to a simplest possible case, that is, two mutually interacting
qubits, immersed in a common bosonic bath. We analyze the reduced
density matrix of one of the two qubits when the state of the
other qubit is regularly projected to one and the same state,
according to the spirit of the purification
scheme~\cite{ref:Nakazato_PRL}. It is shown in the next section
that if the bosonic environment causes only dephasing effects on
the qubits, the purification scheme is robust enough to ensure one
to extract a pure state, provided an appropriate state is
confirmed on the other qubit at suitable intervals, just as in the
ideal case. On the other hand, if the environment brings about a
dissipative effect on the two-qubit system, it is quite difficult
to solve the projected reduced dynamics in a compact form.
However, in Sec.~\ref{sec_dissipation} it is shown that one can
still estimate the behavior of the state analytically for a large,
but finite number of measurements $N$. After an explicit
expression of the projected density matrix for one of the qubits
is derived in a closed form in Sec.~\ref{ssec_prdm}, its
asymptotic behavior for large $N$ is presented in
Sec.~\ref{ssec_asymp}\@. Even under a dissipative environment, one
can find a possibility of extracting a dominant pure state when
the number of measurements $N$ is large, but not extremely large,
provided the dissipation is weak enough. However, it is to be
noted that in the limit of infinite number of measurements, only a
mixed state, except a trivial pure (ground) state, can survive, in
accordance with the general theorem shown in~\cite{ref:PrevPaper}.
We close the paper by giving the summary and discussions in the
final Sec.~\ref{sec_final}\@. A few Appendices are added to
clarify some details that are not shown in the text.

\section{Purification under dephasing environment}
\label{sec_dephasing} Consider a quantum system composed of two
mutually interacting two-level systems (qubits), X and S, immersed
in a common bosonic bath. Assume that the dynamics is given by the
total Hamiltonian of the form
\begin{equation}
H_{\rm T}=H_{\rm XS}+H_{\rm bath}+H_{\rm int},
\end{equation}
where Hamiltonians for the two-qubit system and for the bosonic bath read
\begin{align}
H_{\rm XS}&={\omega\over2}\sigma_z^{({\rm X})}+{\Omega\over2}\sigma_z^{({\rm S})}
+g\,\Bigl(
\sigma_+^{({\rm X})}\sigma_-^{({\rm S})}+\sigma_-^{({\rm X})}\sigma_+^{({\rm S})}
\Bigr),\label{eq:HXS}\\
H_{\rm bath}&=\int d\bm{k}\,\omega_{\bm{k}}a^\dagger_{\bm{k}}a_{\bm{k}}, \label{eq:Hbath}
\end{align}
and the interaction Hamiltonian
\begin{equation}
H_{\rm int}=\Sigma_z\int d\bm{k}\,\Bigl(h(\bm{k})a_{\bm{k}}+h.c.\Bigr)
\end{equation}
causes dephasing on the system.
Here the notations are standard, $\sigma_\pm^{({\rm i})}=(\sigma_x^{({\rm i})}\pm i\sigma_y^{({\rm i})})/2$ ($\text{i}=\text{X}$, S) etc.\ and $\bm{\Sigma}=\bm{\sigma}^{({\rm X})}+\bm{\sigma}^{({\rm S})}$, and the interactions with the bath are assumed to be symmetric between X and S and are given by the same function $h(\bm{k})$, for simplicity.
Since the above Hamiltonian $H_{\rm XS}$ is diagonalized,
\begin{equation}
H_{\rm XS}=E_2|2\rangle\langle2|+E_0|0\rangle\langle0|+E_+|+\rangle\langle+|+E_-|-\rangle\langle-|
\end{equation}
where
\begin{equation}
\begin{matrix}
&|2\rangle=|{\uparrow}\rangle_{\rm X}\otimes|{\uparrow}\rangle_{\rm S}\equiv|{\uparrow\uparrow}\rangle,\hfill\\
\noalign{\smallskip}
&|0\rangle=|{\downarrow\downarrow}\rangle,\hfill\\
\noalign{\smallskip}
&\displaystyle
|\pm\rangle=\frac{1}{\sqrt{2}}\,\Bigl(
\sqrt{1\pm(\omega-\Omega)/2E_+}|{\uparrow\downarrow}\rangle\pm\sqrt{1\mp(\omega-\Omega)/2E_+}|{\downarrow\uparrow}\rangle
\Bigr),
\end{matrix}
\qquad
\begin{matrix}
&E_2=\displaystyle{1\over2}(\omega+\Omega)=-E_0,\hfill\\
\noalign{\medskip}
&\displaystyle
E_+=\sqrt{\frac{1}{4}(\omega-\Omega)^2+g^2}=-E_-,
\end{matrix}
\label{eq:20pm}
\end{equation}
and the interaction is rewritten in terms of these eigenstates as
\begin{equation}
H_{\rm int}=\Bigl(
|2\rangle\langle2|-|0\rangle\langle0|
\Bigr)\int d\bm{k}\,\Bigl(
2h(\bm{k})a_{\bm{k}}+h.c.
\Bigr),
\end{equation}
the usual procedure~\cite{ref:m_eq} yields the master equation for the system
\begin{equation}\label{eq:me_deph}
\dot{\rho}_{\rm XS}=-i[\tilde{H}_{\rm XS},\,\rho_{\rm XS}]
+\gamma\Bigl[\Bigl(|2\rangle\langle2|-|0\rangle\langle0|\Bigr)\rho_{\rm XS}\Bigl(|2\rangle\langle2|-|0\rangle\langle0|\Bigr)-{1\over2}\Bigl\{|2\rangle\langle2|+|0\rangle\langle0|,\,\rho_{\rm XS}\Bigr\}\Bigr],
\end{equation}
with the Hamiltonian $\tilde{H}_{\rm XS}$, which is still
diagonalized as in (\ref{eq:20pm}) but with an energy shift $\Delta E$,
\begin{equation}
\tilde{E}_2=E_2+\Delta E,\qquad
\tilde{E}_0=E_0+\Delta E,\qquad
\tilde{E}_\pm=E_\pm.
\end{equation}
The (dephasing) constant $\gamma$ and the energy shift $\Delta E$
are given by the spectral density as usual~\cite{ref:m_eq}. A
remarkable point is that the structure of the master equation is
not affected by the temperature as in the dissipative case (See
Sec.~\ref{sec_dissipation}), and the temperature only influences
the values of the decay constant and of the energy shift. The
master equation (\ref{eq:me_deph}) is solved~\cite{ref:sol_m_eq}
in the following form~\cite{ref:Kraus}
\begin{equation}
\rho_{\rm XS}(t)=\sum_{i=0,\pm}K_i\,\rho_{\rm XS}(0)K_i^\dagger,
\end{equation}
where the three operators are given by
\begin{align}
&K_0=e^{-i\tilde{E}_2t-\gamma t/2}|2\rangle\langle2|+e^{-i\tilde{E}_0t-\gamma t/2}|0\rangle\langle0|+e^{-iE_+t}|+\rangle\langle+|+e^{-iE_-t}|-\rangle\langle-|,\nonumber\\
&K_+=\sqrt{\cosh\gamma t-1}\,\Bigl(
e^{-i\tilde{E}_2t-\gamma t/2}|2\rangle\langle2|+e^{-i\tilde{E}_0t-\gamma t/2}|0\rangle\langle0|
\Bigr),\nonumber\\
&K_-=\sqrt{\sinh\gamma t}\,\Bigl(
e^{-i\tilde{E}_2t-\gamma t/2}|2\rangle\langle2|-e^{-i\tilde{E}_0t-\gamma t/2}|0\rangle\langle0|
\Bigr).
\end{align}
Observe that these operators are all diagonal and the environment
causes no transitions between different levels. Only the phases
between different levels are affected by this type of interaction
with the environment and this is the reason why it is called a
dephasing interaction.

According to the spirit of the repeated-measurement-based
purification scheme~\cite{ref:Nakazato_PRL}, we measure qubit X to
confirm that it is in the same, particular state, while we do not
touch the other qubit S, even though it is affected indirectly
through its interaction with X\@. This process is repeated many
times at regular intervals and we are interested in what happens
to the state of qubit S, which also suffers the dephasing effect
from the bath in addition to the projective actions on X\@. It has
been shown in \cite{ref:PrevPaper} that for a pure state to be
finally extracted in this kind of process with projections and
decoherence, it has to be one of the simultaneous eigenstates of
all the relevant projected operators and furthermore its
eigenvalue for the projected dynamics has to be the largest in
magnitude, preferably close to unity. In our case of dephasing
environment, it is easy to find such simultaneous eigenstates of
the projected operators, as will be shown below.

Let us consider to measure the up state $|{\uparrow}\rangle_{\rm
X}$ regularly at $t=n\tau$ ($n=1,2,\ldots$), since we already know
that this type of measurement results in an optimal purification
of qubit S in the ideal case~\cite{ref:p_appl}. The relevant
projected operators $V_{\uparrow(i)}={}_{\rm
X}\langle{\uparrow}\nobreak|K_i|{\uparrow}\rangle_{\rm X}$ read
(the subscript ${}_{\rm S}$ shall be suppressed in the following)
\begin{align}
&V_{\uparrow(0)}={}_{\rm X}\langle{\uparrow}|K_0|{\uparrow}\rangle_{\rm X}=e^{-i\tilde{E}_2\tau-\gamma\tau/2}|{\uparrow}\rangle\langle{\uparrow}|+
\left(
\cos E_+\tau+i{\omega-\Omega\over2E_+}\sin E_+\tau
\right)|{\downarrow}\rangle\langle{\downarrow}|,\nonumber\\
&V_{\uparrow(+)}={}_{\rm X}\langle{\uparrow}|K_+|{\uparrow}\rangle_{\rm X}=\sqrt{\cosh\gamma\tau-1}\,e^{-i\tilde{E}_2\tau-\gamma\tau/2}|{\uparrow}\rangle\langle{\uparrow}|,\nonumber\\
&V_{\uparrow(-)}={}_{\rm X}\langle{\uparrow}|K_-|{\uparrow}\rangle_{\rm X}=\sqrt{\sinh\gamma\tau}\,e^{-i\tilde{E}_2\tau-\gamma\tau/2}|{\uparrow}\rangle\langle{\uparrow}|.
\end{align}
It is clear that both the up and down states (of qubit S) are the
simultaneous eigenstates of all the above projected operators and
the eigenvalues and eigenstates for the projected dynamics, which
is given by the following map
\begin{equation}
\rho(k\tau)
=\sum_{i=0,\pm}V_{\uparrow(i)}\rho\bm{(}(k-1)\tau\bm{)}V_{\uparrow(i)}^\dagger,
\end{equation}
are read from
\begin{align}
\sum_{i=0,\pm}V_{\uparrow(i)}\,|{\uparrow}\rangle\langle{\uparrow}|V_{\uparrow(i)}^\dagger
&=\left(
\bigl|e^{-i\tilde{E}_2\tau-\gamma\tau/2}\bigr|^2+\bigl|\sqrt{\cosh\gamma\tau-1}\,e^{-i\tilde{E}_2\tau-\gamma\tau/2}\bigr|^2+\bigl|\sqrt{\sinh\gamma\tau}\,e^{-i\tilde{E}_2\tau-\gamma\tau/2}\bigr|^2
\right)|{\uparrow}\rangle\langle{\uparrow}|\label{eq:VupupV}\nonumber\\
&=|{\uparrow}\rangle\langle{\uparrow}|,\\
\sum_{i=0,\pm}V_{\uparrow(i)}\,|{\downarrow}\rangle\langle{\downarrow}|V_{\uparrow(i)}^\dagger
&=\left|\cos E_+\tau+i{\omega-\Omega\over2E_+}\sin
E_+\tau\right|^2|{\downarrow}\rangle\langle{\downarrow}|
\equiv|\xi|^2|{\downarrow}\rangle\langle{\downarrow}|
,\\
\sum_{i=0,\pm}V_{\uparrow(i)}\,|{\uparrow}\rangle\langle{\downarrow}|V_{\uparrow(i)}^\dagger
&=e^{-i\tilde{E}_2\tau-\gamma\tau/2}
\left(
\cos
E_+\tau-i{\omega-\Omega\over2E_+}\sin
E_+\tau
\right)|{\uparrow}\rangle\langle{\downarrow}|
=e^{-i\tilde{E}_2\tau-\gamma\tau/2}\xi^*|{\uparrow}\rangle\langle{\downarrow}|
,\\
\sum_{i=0,\pm}V_{\uparrow(i)}\,|{\downarrow}\rangle\langle{\uparrow}|V_{\uparrow(i)}^\dagger
&=e^{i\tilde{E}_2\tau-\gamma\tau/2}
\left(
\cos
E_+\tau+i{\omega-\Omega\over2E_+}\sin
E_+\tau
\right)|{\downarrow}\rangle\langle{\uparrow}|
=e^{i\tilde{E}_2\tau-\gamma\tau/2}\xi|{\downarrow}\rangle\langle{\uparrow}|.
\end{align}

Starting from a factorized initial state,
$\rho_{\rm XS}(0)=|{\uparrow}\rangle_\text{X}\langle{\uparrow}|\otimes\rho(0)$ \footnote{This assumption is not restrictive, since we can always prepare the total system in this class of states by performing a measurement on X to confirm it in the state $|{\uparrow}\rangle_\text{X}$.}, the density matrix is given
by
\begin{equation}
\rho(N\tau) =\begin{pmatrix}
 \rho_{\uparrow\uparrow}(0)
 &\left(e^{-i\tilde{E}_2\tau-\gamma\tau/2}\xi^*\right)^{N}\rho_{\uparrow\downarrow}(0)\\
 \noalign{\smallskip}
 \left(e^{i\tilde{E}_2\tau-\gamma\tau/2}\xi\right)^{N}\rho_{\downarrow\uparrow}(0)&
 |\xi|^{2 N}\,\rho_{\downarrow\downarrow}(0)
 \end{pmatrix}.
\end{equation}
Observe that the up state $|{\uparrow}\rangle\langle{\uparrow}|$
is the eigenstate belonging to the eigenvalue unity and that there
are no $\gamma$'s left in its relation~(\ref{eq:VupupV}). It means
that the dephasing has no effect on the ability of the
purification scheme. The other eigenstates belong to eigenvalues
smaller than unity if $\sin E_+\tau\not=0$, just as in the ideal
case with no decoherence. The above relations clearly show that we
will be able to purify qubit S to the up state with no loss of
probability, unless $\cos E_+\tau=\pm 1$, irrespective of its
initial mixed state, when the qubit X is repeatedly confirmed to
be in the up state. The repeated-measurement-based purification
scheme is thus shown to be robust enough against the dephasing
effect, at least when $|{\uparrow}\rangle_{\rm X}$ is measured. On
the other hand, it is easy to demonstrate that the only two
possible pure states that can be extracted under dephasing
environment are $|{\uparrow}\rangle$ and $|{\downarrow}\rangle$.
Indeed, the projected operators ${}_{\rm
X}\langle\phi|K_\pm|\phi\rangle_{\rm X},\,\forall\phi$ are always
diagonal in the basis $\{|{\uparrow}\rangle,\,
|{\downarrow}\rangle\}$, and the operator $K_0$ admits one of such
two states as an eigenstate only when $|{\uparrow}\rangle_{\rm X}$
or $|{\downarrow}\rangle_{\rm X}$ is measured. It is worth
stressing the possibility of extracting a pure state, i.e.
$|{\uparrow}\rangle$, even under dephasing in certain conditions.

\section{Purification under dissipative environment}
\label{sec_dissipation}
Consider next a dissipative environment and assume that it interacts with the two qubits, X and S, through the interaction Hamiltonian
\begin{equation}
H_{\rm int}=\int d\bm{k}\,h(\bm{k})(\Sigma_+a_{\bm{k}}+\Sigma_-a^\dagger_{\bm{k}}),\qquad
\Sigma_\pm=\sigma_\pm^{(\rm X)}+\sigma_\pm^{(\rm S)}.
\label{eq:HIdiss}
\end{equation}
For the sake of simplicity, we shall set $\omega=\Omega$
\footnote{The condition $\Omega=\omega$ is readily and naturally
follows, if, for instance, the two qubits are realized by two
spins with the same magnetic moment. Another example would be
given by two identical atoms in a linear trap.}, though
generalization would be straightforward. The system Hamiltonian
reads
\begin{equation}
H_{\rm XS}={\Omega\over2}\Sigma_3+g\left(\sigma_+^{({\rm X})}\sigma_-^{({\rm S})}+\sigma_-^{({\rm X})}\sigma_+^{({\rm S})}\right)
=\sum_{i=2,0,\pm}E_i\,|i\rangle\langle i|,
\end{equation}
where eigenenergies and eigenstates are somewhat simplified
\begin{equation}
\begin{matrix}
&|2\rangle
=|{\uparrow\uparrow}\rangle,\hfill\\
\noalign{\smallskip}
&|0\rangle=|{\downarrow\downarrow}\rangle,\hfill\\
\noalign{\smallskip}
&\displaystyle
|\pm\rangle=\frac{1}{\sqrt{2}}\,\Bigl(
|{\uparrow\downarrow}\rangle\pm|{\downarrow\uparrow}\rangle
\Bigr),\hfill\\
\end{matrix}
\qquad
\begin{matrix}
&E_2=\Omega=-E_0,\hfill\\
\noalign{\medskip}
&E_\pm=\pm g\hfill
\end{matrix}
\label{eq:20pm_diss}
\end{equation}
and the interaction Hamiltonian is rewritten as
\begin{equation}
H_{\rm int}=\int d\bm{k}\,\sqrt2\,h(\bm{k})\left[\Bigl(
|0\rangle\langle+|+|+\rangle\langle2|
\Bigr)\,a^\dagger_{\bm{k}}+h.c.\right].\label{eq:Hint}
\end{equation}
When the bosonic bath is at temperature $T=0$, the master equation is derived under the usual conditions~\cite{ref:m_eq}
\begin{equation}
\dot\rho_{\rm XS}
=-i[\tilde{H}_{\rm XS},\,\rho_{\rm XS}]
 +\gamma\Bigl[|+\rangle\langle2|\rho_{\rm XS}|2\rangle\langle+|
 -{1\over2}\Bigl\{|2\rangle\langle2|,\,\rho_{\rm XS}\Bigr\}\Bigr]
+\gamma\Bigl[|0\rangle\langle+|\rho_{\rm XS}|+\rangle\langle0|
 -{1\over2}\Bigl\{|+\rangle\langle+|,\,\rho_{\rm XS}\Bigr\}\Bigr].\label{eq:m_eq2}
\end{equation}
The decay constants $\gamma_{2\to+}$ and $\gamma_{+\to0}$ (we
assume $\Omega>g$), given by the on-shell form factors, are
assumed to be the same
$\gamma\equiv\gamma_{2\to+}=\gamma_{+\to0}$, just for simplicity.
The Hamiltonian $\tilde{H}_{\rm XS}$ is still diagonalized by
$|i\rangle$ ($i=2,0,\pm$) in (\ref{eq:20pm_diss})
\begin{equation}
\tilde{H}_{\rm XS}=\sum_{i=2,0,\pm}\tilde{E}_i\,|i\rangle\langle i|,\qquad
\tilde{E}_2=E_2+\Delta E_2,\quad
\tilde{E}_+=E_++\Delta E_+,\quad
\tilde{E}_0=E_0,\quad
\tilde{E}_-=E_-.
\end{equation}

We follow the recently developed technique~\cite{ref:sol_m_eq} to obtain the solution of the master equation (\ref{eq:m_eq2}) in the following compact form
\begin{equation}
\rho_{\rm XS}(t)
=e^{At}\rho_{\rm XS}(0)e^{A^\dagger t}
+(1-e^{-\gamma t})B_0\rho_{\rm XS}(0){B_0}^\dagger
+(1-e^{-\gamma t}-\gamma te^{-\gamma t})B_0B_1\rho_{\rm XS}(0){B_1}^\dagger{B_0}^\dagger
+\gamma te^{-\gamma t}B_1\rho_{\rm XS}(0){B_1}^\dagger,
\label{eq:sol1}
\end{equation}
where
\begin{equation}
e^{At}=e^{-i\tilde{E}_2t-\gamma t/2}|2\rangle\langle2|
+e^{-i\tilde{E}_+t-\gamma t/2}|+\rangle\langle+|
+e^{-iE_0t}|0\rangle\langle0|
+e^{-iE_-t}|-\rangle\langle-|
\end{equation}
and
\begin{equation}
B_0=|0\rangle\langle+|,\qquad B_1=|+\rangle\langle2|.
\end{equation}

Notice that one can think of a similar master equation
that describes the dynamics of the two qubits X and S, in
interaction with a common bosonic bath at temperature $T=0$
through the same coupling (\ref{eq:HIdiss}), but with a different
mutual coupling from (\ref{eq:HXS}). For our purpose, however, it
is desirable to make things as simple as possible, for, as will be
seen below, even the above seemingly simplified dynamics
(\ref{eq:m_eq2}) can bring us with quite involved expressions for
the reduced density matrix of S when the state of X is
periodically projected on one and the same state. Therefore, in
this paper, we exclusively consider the dynamics (\ref{eq:m_eq2})
and endeavor to disclose the asymptotic behavior of the projected
reduced density matrix for system S\@.

If the system X is to be repeatedly measured at $t=n\tau$
($n=1,2,\ldots$) to confirm that it is in the state
$|\alpha\rangle_{\rm X}=\alpha|{\uparrow}\rangle_{\rm
X}+|{\downarrow}\rangle_{\rm X}$ (the normalization factor
$(1+|\alpha|^2)^{-1/2}$ is tentatively omitted here for notational
simplicity), the relevant operators for the dynamics of qubit S
read
\begin{align}
&{}_{\rm X}\langle\alpha|e^{A\tau}|\alpha\rangle_{\rm X}
=e^{-i\tilde{E}_2\tau-\gamma \tau/2}|\alpha|^2|{\uparrow}\rangle\langle{\uparrow}|
+{1\over2}e^{-i\tilde{E}_+\tau-\gamma \tau/2}
\Bigl(\alpha^*|{\downarrow}\rangle+|{\uparrow}\rangle\Bigr)\Bigl(\alpha\langle{\downarrow}|+\langle{\uparrow}|\Bigr)\nonumber\\
&\phantom{{}_{\rm X}\langle\alpha|e^{A\tau}|\alpha\rangle_{\rm X}
=}
+e^{-iE_0\tau}|{\downarrow}\rangle\langle{\downarrow}|
+{1\over2}e^{-iE_-\tau}\Bigl(\alpha^*|{\downarrow}\rangle-|{\uparrow}\rangle\Bigr)\Bigl(\alpha\langle{\downarrow}|-\langle{\uparrow}|\Bigr)\nonumber\\
\noalign{\medskip}
&\phantom{{}_{\rm X}\langle\alpha|e^{A\tau}|\alpha\rangle_{\rm X}
}
=\left[\begin{matrix}\displaystyle
|\alpha|^2e^{-i\tilde{E}_2\tau-\gamma \tau/2}+{1\over2}e^{-i\tilde{E}_+\tau-\gamma \tau/2}+{1\over2}e^{-iE_-\tau}
&\displaystyle
{\alpha\over2}(e^{-i\tilde{E}_+\tau-\gamma \tau/2}-e^{-iE_-\tau})\\
\noalign{\medskip}
\displaystyle
{\alpha^*\over2}(e^{-i\tilde{E}_+\tau-\gamma \tau/2}-e^{-iE_-\tau})\quad
&\displaystyle
\quad{|\alpha|^2\over2}(e^{-i\tilde{E}_+\tau-\gamma \tau/2}+e^{-iE_-\tau})+e^{-iE_0\tau}
 \end{matrix}\right]\nonumber\\
\noalign{\medskip}
&\phantom{{}_{\rm X}\langle\alpha|e^{A\tau}|\alpha\rangle_{\rm X}
}
\equiv V,\\
\noalign{\medskip}
&{}_{\rm X}\langle\alpha|B_0|\alpha\rangle_{\rm X}
={1\over\sqrt2}|{\downarrow}\rangle\Bigl(\alpha\langle{\downarrow}|+\langle{\uparrow}|\Bigr)
=\left[
\begin{matrix}
0&0\\
\noalign{\medskip}
\displaystyle{1\over\sqrt2}\quad
&\displaystyle\quad{\alpha\over\sqrt2}
\end{matrix}\right]\equiv C_0,\\
\noalign{\medskip}
&{}_{\rm X}\langle\alpha|B_0B_1|\alpha\rangle_{\rm X}
=\alpha|{\downarrow}\rangle\langle{\uparrow}|
=\left[
\begin{matrix}
0&0\\
\noalign{\medskip}
\alpha
&0
\end{matrix}\right]\equiv C_1,\\
\noalign{\medskip}
&{}_{\rm X}\langle\alpha|B_1|\alpha\rangle_{\rm X}
={\alpha\over\sqrt2}\Bigl(\alpha^*|{\downarrow}\rangle+|{\uparrow}\rangle\Bigr)\langle{\uparrow}|
=\left[
\begin{matrix}
\displaystyle{\alpha\over\sqrt2}&0\\
\noalign{\medskip}
\displaystyle{|\alpha|^2\over\sqrt2}
&0
\end{matrix}\right]\equiv C_2.
\end{align}
The projected reduced density matrix $\rho(\tau)$ of qubit S is given by the following map
\begin{equation}
\rho(\tau)=V\rho(0)V^\dagger
+(1-e^{-\gamma \tau})C_0\rho(0){C_0}^\dagger
+(1-e^{-\gamma \tau}-\gamma\tau e^{-\gamma \tau})C_1\rho(0){C_1}^\dagger
+\gamma\tau e^{-\gamma \tau}C_2\rho(0){C_2}^\dagger.
\label{eq:rho(tau)}
\end{equation}

\subsection{Projected reduced density matrix}\label{ssec_prdm}
It is not difficult to confirm that the down state
$|{\downarrow}\rangle$ can be a simultaneous eigenstate of all the
above relevant operators if $\alpha=0$. This corresponds to the
case where qubit X is confirmed to be in the down state and qubit
S is projected to the down state
$|{\downarrow}\rangle\langle{\downarrow}|$ with probability 1.
This fact has already been
pointed out in Ref.~\cite{ref:PrevPaper}. Furthermore, one can
easily see that for nonvanishing $\alpha$, these relevant
operators do not have a common eigenstate and therefore no pure
state can be extracted in the limit of infinite number of
measurements, according to the general
theorem~\cite{ref:PrevPaper}. Nevertheless we are interested in
the asymptotic behavior of the projected reduced density matrix
$\rho(N\tau)$ after $N$ successive measurements, in the hope of
finding a way out of, or bypassing such a ``no-go'' theorem.

In order to find the asymptotic form of the projected dynamics when the projection is repeated many times, observe first that the above operators $C_0,\,C_1$, which are expressed as (now the normalization constant is recovered)
\begin{equation}
C_0={1\over\sqrt2(1+|\alpha|^2)}\begin{pmatrix}0\\1\end{pmatrix}(1,\alpha),\qquad
C_1={\alpha\over1+|\alpha|^2}\begin{pmatrix}0\\1\end{pmatrix}(1,0)
\end{equation}
in the up-down basis, enforce the system to be in a pure (down) state
\begin{equation}
\begin{pmatrix}0\\1\end{pmatrix}(0,1)=|{\downarrow}\rangle\langle{\downarrow}|
\equiv\rho_{\downarrow\downarrow},
\end{equation}
while the operator $C_2$, which can also be expressed as
\begin{equation}
C_2={\alpha\over\sqrt2(1+|\alpha|^2)}\begin{pmatrix}1\\\alpha^*\end{pmatrix}(1,0),
\end{equation}
drives the system to another pure state
\begin{equation}
{1\over1+|\alpha|^2}\begin{pmatrix}1\\\alpha^*\end{pmatrix}(1,\alpha)
\equiv|\alpha^*\rangle\langle\alpha^*|\equiv\rho_{\alpha^*},
\end{equation}
irrespectively of the state the system had lived in just before projection.
This means that in general the projected dynamics drives the system to a mixed state, which would be made more apparent if it is written in the following form (apart from normalization)
\begin{equation}\label{rho1}
\rho(\tau)
=V\rho(0)V^\dagger
+F\bm{(}\rho(0)\bm{)}\rho_{\downarrow\downarrow}
+G\bm{(}\rho(0)\bm{)}\rho_{\alpha^*},
\end{equation}
where the positive constants $F$ and $G$ read
\begin{equation}
F\bm{(}\rho(0)\bm{)}
={1-e^{-\gamma \tau}\over2(1+|\alpha|^2)}\langle\alpha^*|\rho(0)|\alpha^*\rangle
+{|\alpha|^2\over(1+|\alpha|^2)^2}(1-e^{-\gamma \tau}-\gamma \tau e^{-\gamma \tau})\langle{\uparrow}|\rho(0)|{\uparrow}\rangle
\end{equation}
and
\begin{equation}
G\bm{(}\rho(0)\bm{)}
={|\alpha|^2\over2(1+|\alpha|^2)}\gamma \tau e^{-\gamma \tau}\langle{\uparrow}|\rho(0)|{\uparrow}\rangle,
\end{equation}
respectively.

It is not difficult to see that after $N$ repetitions of the projected dynamics, the state of the system S, apart from the normalization, is driven to [$\rho_0\equiv\rho(0)$]
\begin{equation}\label{rhoN}
\rho_N\equiv\rho(N\tau)=V^N\rho(0)V^\dagger{}^N
+\sum_{k=0}^{N-1}[
F(\rho_k)V^{N-1-k}\rho_{\downarrow\downarrow}V^\dagger{}^{N-1-k}
+G(\rho_k)V^{N-1-k}\rho_{\alpha^*}V^\dagger{}^{N-1-k}
].
\end{equation}
The coefficient $F(\rho_N)$ satisfies
\begin{align}
F(\rho_N)&={1-e^{-\gamma \tau}\over2(1+|\alpha|^2)}\langle\alpha^*|\rho_N|\alpha^*\rangle
+{|\alpha|^2(1-e^{-\gamma \tau}-\gamma \tau e^{-\gamma \tau})\over(1+|\alpha|^2)^2}\langle{\uparrow}|\rho_N|{\uparrow}\rangle\nonumber\\
&={1-e^{-\gamma \tau}\over2(1+|\alpha|^2)}\langle\alpha^*|V^N\rho(0)V^\dagger{}^N|\alpha^*\rangle
+{|\alpha|^2(1-e^{-\gamma \tau}-\gamma \tau e^{-\gamma \tau})\over(1+|\alpha|^2)^2}\langle{\uparrow}|V^N\rho(0)V^\dagger{}^N|{\uparrow}\rangle\nonumber\\
&\qquad+\sum_{k=0}^{N-1}F(\rho_k)
\left[{1-e^{-\gamma \tau}\over2(1+|\alpha|^2)}\Bigl|\langle\alpha^*|V^{N-1-k}|{\downarrow}\rangle\Bigr|^2+{|\alpha|^2(1-e^{-\gamma \tau}-\gamma \tau e^{-\gamma \tau})\over(1+|\alpha|^2)^2}\Bigl|\langle{\uparrow}|V^{N-1-k}|{\downarrow}\rangle\Bigr|^2\right]
\nonumber\\
&\qquad+\sum_{k=0}^{N-1}G(\rho_k)
\left[{1-e^{-\gamma \tau}\over2(1+|\alpha|^2)}\Bigl|\langle\alpha^*|V^{N-1-k}|\alpha^*\rangle\Bigr|^2+{|\alpha|^2(1-e^{-\gamma \tau}-\gamma \tau e^{-\gamma \tau})\over(1+|\alpha|^2)^2}\Bigl|\langle{\uparrow}|V^{N-1-k}|\alpha^*\rangle\Bigr|^2\right].
\end{align}
Similarly, we have
\begin{align}
G(\rho_N)&={|\alpha|^2\gamma \tau e^{-\gamma \tau}\over2(1+|\alpha|^2)}\langle{\uparrow}|\rho_N|{\uparrow}\rangle\nonumber\\
&={|\alpha|^2\gamma \tau e^{-\gamma \tau}\over2(1+|\alpha|^2)}\langle{\uparrow}|V^N\rho(0)V^\dagger{}^N|{\uparrow}\rangle\nonumber\\
&\qquad+\sum_{k=0}^{N-1}F(\rho_k){|\alpha|^2\gamma \tau e^{-\gamma \tau}\over2(1+|\alpha|^2)}
\Bigl|\langle{\uparrow}|V^{N-1-k}|{\downarrow}\rangle\Bigr|^2
+\sum_{k=0}^{N-1}G(\rho_k){|\alpha|^2\gamma \tau e^{-\gamma \tau}\over2(1+|\alpha|^2)}
\Bigl|\langle{\uparrow}|V^{N-1-k}|\alpha^*\rangle\Bigr|^2.
\end{align}
It is now clear that these coefficients satisfy the recursion relation of the following form
\begin{equation}
\left[
\begin{matrix}
F(\rho_N)\\\noalign{\smallskip}G(\rho_N)\end{matrix}
\right]
=\sum_{k=0}^{N-1}{\cal A}_{N-1-k}
\left[
\begin{matrix}
F(\rho_k)\\\noalign{\smallskip}G(\rho_k)\end{matrix}
\right]
+b_N,
\end{equation}
where two-by-two matrices ${\cal A}_{N-1-k}$ and two-component column vectors $b_N$ would be evident from the previous expressions.
If a parameter $x$ is introduced, its formal solution is easily found through an identity relation
\begin{equation}\label{id_relation}
\sum_{n=0}^\infty x^n\left[\begin{matrix}F(\rho_n)\\\noalign{\smallskip}G(\rho_n)\end{matrix}
\right]=\left(1-x\sum_{k=0}^\infty x^k{\cal A}_k\right)^{-1}\sum_{n=0}^\infty x^nb_n,
\end{equation}
that is, the coefficients $F(\rho_k)$ and $G(\rho_k)$ are given as those of the $k$th power $x^k$ on the right hand side.
Plugging these values into (\ref{rhoN}), we obtain the explicit expression of the density matrix of qubit S at $N\tau$.

We assume the diagonalizability of $V$ and introduce the
eigenvectors of $V$
\begin{equation}
V|u_i\rangle=\lambda_i|u_i\rangle,\quad
\langle v_i|V=\lambda_i\langle v_i|,\quad
\sum_{i=1}^2|u_i\rangle\langle v_i|=1,\quad
\langle v_i|u_j\rangle=\delta_{ij},
\end{equation}
to expand it as
\begin{equation}
V=\sum_{i=1}^2\lambda_i|u_i\rangle\langle v_i|.
\end{equation}
The elements of the two-by-two matrix ${\cal A}_k$ and the column vector $b_k$ are then expressed as
\begin{equation}
({\cal A}_k)_{ij}=\sum_{a,b}(\Lambda_{ab})^k({\cal C}^{ab})_{ij},\quad(b_k)_i=\sum_{a,b}(\Lambda_{ab})^k(d^{ab})_i,
\end{equation}
where
\begin{equation}
\Lambda_{ab}\equiv\lambda_a\lambda_b^*
\end{equation}
and the explicit expressions of $({\cal C}^{ab})_{ij}$ and $(d^{ab})_i$ are found in Appendix~\ref{app_Cabdab}.
The above identity relation (\ref{id_relation}) tells us that the coefficients $F(\rho_k)$ and $G(\rho_k)$ are explicitly given by
\begin{align}\label{FandG0}
&\left[
\begin{matrix}
F(\rho_k)\\\noalign{\smallskip}G(\rho_k)\end{matrix}
\right]
=\sum_{{0\le\ell,k_1,\ldots,k_\ell,n\le k\atop
        \ell+k_1+\cdots+k_\ell+n=k}}
 {\cal A}_{k_1}{\cal A}_{k_2}\cdots{\cal A}_{k_\ell}b_n\nonumber\\
&\quad=\sum_{\ell=0}^k\sum_{k_1=0}^{k-\ell}\sum_{k_2=0}^{k-\ell-k_1}\cdots
\sum_{k_\ell=0}^{k-\ell-k_1-\cdots-k_{\ell-1}}\sum_{a_i,b_j\atop 1\le i,j\le\ell+1}
(\Lambda_{a_1b_1})^{k_1}\cdots(\Lambda_{a_\ell b_\ell})^{k_\ell}(\Lambda_{a_{\ell+1}b_{\ell+1}})^{k-\ell-k_1-\cdots-k_\ell}
{\cal C}^{a_1b_1}\cdots{\cal C}^{a_\ell b_\ell}d^{a_{\ell+1}b_{\ell+1}}.
\end{align}
Observe that the indices $k_1,\ldots,k_\ell$ only appear as the exponents of the eigenvalues.
The summations over these indices are not difficult to perform and each summation results in a similar expression.
Actually, since we can prove by induction (see Appendix~\ref{app1}),
\begin{align}
\sum_{k_1=0}^{k-\ell}\sum_{k_2=0}^{k-\ell-k_1}\cdots
\sum_{k_\ell=0}^{k-\ell-k_1-\cdots-k_{\ell-1}}x_1^{k_1}\cdots x_\ell^{k_\ell}x_{\ell+1}^{k-\ell-k_1-\cdots-k_\ell}
=&\,{x_{\ell+1}^k\over(x_{\ell+1}-x_\ell)(x_{\ell+1}-x_{\ell-1})\cdots(x_{\ell+1}-x_1)}\nonumber\\
&+{x_\ell^k\over(x_\ell-x_{\ell+1})(x_\ell-x_{\ell-1})\cdots(x_\ell-x_1)}\nonumber\\
&{}+\cdots+{x_1^k\over(x_1-x_{\ell+1})(x_1-x_\ell)\cdots(x_1-x_2)}\nonumber\\
=&\,\sum_{m=1}^{\ell+1}{x_m^k\over\prod\limits_{{n\not=m\atop1\le n\le\ell+1}}(x_m-x_n)},
\label{eq:relation1}
\end{align}
we arrive at
\begin{equation}\label{FandG}
\left[
\begin{matrix}
F(\rho_k)\\\noalign{\smallskip}G(\rho_k)\end{matrix}
\right]
=\sum_{\ell=0}^k\sum_{m=1}^{\ell+1}\sum_{a_\alpha,b_\beta\atop1\le\alpha,\beta\le\ell+1}
{(\Lambda_{a_mb_m})^k\over\prod\limits_{n\not=m\atop1\le n\le\ell+1}(\Lambda_{a_mb_m}-\Lambda_{a_nb_n})}\,
{\cal C}^{a_1b_1}\cdots{\cal C}^{a_\ell b_\ell}d^{a_{\ell+1}b_{\ell+1}}.
\end{equation}
Notice that $\rho_N$ in (\ref{rhoN}) is expressed in terms of the eigenvectors $|u_i\rangle$ and $|v_j\rangle$ as
\begin{equation}\label{rhoN2}
\rho_N=\sum_{i,j}(\Lambda_{ij})^N|u_i\rangle\langle v_i|\rho(0)|v_j\rangle\langle u_j|
+\sum_{i,j}\sum_{k=0}^{N-1}(\Lambda_{ij})^{N-1-k}|u_i\rangle\langle u_j|
\Bigl(\langle v_i|\rho_{\downarrow\downarrow}|v_j\rangle,\,\langle v_i|\rho_{\alpha^*}|v_j\rangle\Bigr)
\left[
\begin{matrix}
F(\rho_k)\\\noalign{\smallskip}G(\rho_k)\end{matrix}
\right].
\end{equation}
Plugging the explicit forms of $F(\rho_k)$ and $G(\rho_k)$ (\ref{FandG}) into (\ref{rhoN2}), we are left with the following terms to be evaluated
\begin{align}
&\sum_{k=0}^{N-1}(\Lambda_{ij})^{N-1-k}
\left[
\begin{matrix}
F(\rho_k)\\\noalign{\smallskip}G(\rho_k)\end{matrix}
\right]\nonumber\\
&\quad
=\sum_{k=0}^{N-1}(\Lambda_{ij})^{N-1-k}
\sum_{\ell=0}^k\sum_{m=1}^{\ell+1}\sum_{a_\alpha,b_\beta\atop1\le\alpha,\beta\le\ell+1}
{(\Lambda_{a_mb_m})^k\over\prod\limits_{n\not=m\atop1\le n\le\ell+1}(\Lambda_{a_mb_m}-\Lambda_{a_nb_n})}\,
{\cal C}^{a_1b_1}\cdots{\cal C}^{a_\ell b_\ell}d^{a_{\ell+1}b_{\ell+1}}.\\
\noalign{\noindent
The summations over $k$ and $\ell$ can be interchanged $\sum_{k=0}^{N-1}\sum_{\ell=0}^k=\sum_{\ell=0}^{N-1}\sum_{k=\ell}^{N-1}$ and that over $k$ is performed to yield}
&\quad
=\sum_{\ell=0}^{N-1}\sum_{m=1}^{\ell+1}\sum_{a_\alpha,b_\beta\atop1\le\alpha,\beta\le\ell+1}
{1\over\prod\limits_{n\not=m\atop1\le n\le\ell+1}(\Lambda_{a_mb_m}-\Lambda_{a_nb_n})}
{(\Lambda_{ij})^N(\Lambda_{a_mb_m}/\Lambda_{ij})^\ell-(\Lambda_{a_mb_m})^N\over\Lambda_{ij}-\Lambda_{a_mb_m}}\,
{\cal C}^{a_1b_1}\cdots{\cal C}^{a_\ell b_\ell}d^{a_{\ell+1}b_{\ell+1}}.\label{sumoverk}
\end{align}
This is the explicit form of the second term of the rhs of (\ref{rhoN2}) and brings us with the exact expression of $\rho_N=\rho(N\tau)$.

\subsection{Asymptotic behavior of the projected reduced density matrix}\label{ssec_asymp}
We are now in a position to evaluate the asymptotic form of
$\rho_N$ for large $N$. The particular form seen in
(\ref{sumoverk}) implies that the dominant contributions are due,
in general, to those terms with the same set of indices
$\{ij\}={}\{a_1b_1\}=\{a_2b_2\}=\cdots=\{a_{\ell+1}b_{\ell+1}\}$,
for in such a case there are order-$N$ terms contributing
constructively. This can be understood by looking at the original
expression (\ref{FandG0}). See Appendix~\ref{app_summations} for
more explanations. The dominant contributions to the second term
on the rhs of (\ref{rhoN2})
\begin{align}\label{2ndterm}
\sum_{i,j}\sum_{k=0}^{N-1}(\Lambda_{ij})^{N-1-k}|u_i\rangle\langle u_j|
&\Bigl(\langle v_i|\rho_{\downarrow\downarrow}|v_j\rangle,\,\langle v_i|\rho_{\alpha^*}|v_j\rangle\Bigr)
\left[
\begin{matrix}
F(\rho_k)\\\noalign{\smallskip}G(\rho_k)\end{matrix}\right]\nonumber\\
&=|u_1\rangle\langle u_1|\Bigl(\langle v_1|\rho_{\downarrow\downarrow}|v_1\rangle,\,\langle v_1|\rho_{\alpha^*}|v_1\rangle\Bigr){\cal M}_{11}
+|u_2\rangle\langle u_2|\Bigl(\langle v_2|\rho_{\downarrow\downarrow}|v_2\rangle,\,\langle v_2|\rho_{\alpha^*}|v_2\rangle\Bigr){\cal M}_{22}\nonumber\\
&\quad+|u_1\rangle\langle u_2|\Bigl(\langle v_1|\rho_{\downarrow\downarrow}|v_2\rangle,\,\langle v_1|\rho_{\alpha^*}|v_2\rangle\Bigr){\cal M}_{12}+h.c.
\end{align}
are found in the above (column) vectors ${\cal M}_{ij}$:
\begin{align}\label{M11}
{\cal M}_{11}
\sim{}&\sum_{\ell=0}^{N-1}\Lambda_{11}^{N-1-\ell}{\cal A}_0^\ell
b_0
+\sum_{\ell=0}^{N-1}\sum_{k=\ell+1}^{N-1}\sum_{k_1=0}^{k-\ell}\cdots\sum_{k_\ell=0}^{k-\ell-k_1-\cdots-k_{\ell-1}}\Lambda_{11}^{N-1-\ell}
({\cal C}^{11})^\ell d^{11}
\nonumber\\
={}&(\Lambda_{11}-{\cal A}_0)^{-1}(\Lambda_{11}^N-{\cal A}_0^N)b_0
+\sum_{k=1}^{N-1}\sum_{\ell=0}^{k-1}
{k\choose\ell}
\Lambda_{11}^{N-1-\ell}({\cal C}^{11})^\ell d^{11},
\end{align}
the second term of which is evaluated to be
\begin{align}
&\sum_{k=1}^{N-1}\sum_{\ell=0}^{k-1}{k\choose\ell}
\Lambda_{11}^{N-1-\ell}({\cal C}^{11})^\ell d^{11}
\nonumber\\
&\quad
=-\Lambda_{11}^N({\cal C}^{11})^{-1}\left(1+{{\cal C}^{11}\over\Lambda_{11}}\right)\left[1-\left(1+{{\cal C}^{11}\over\Lambda_{11}}\right)^{N-1}\right]d^{11}
-\Lambda_{11}^{N-2}{\cal C}^{11}\left(1-{{\cal C}^{11}\over\Lambda_{11}}\right)^{-1}\left[1-\left({{\cal C}^{11}\over\Lambda_{11}}\right)^{N-1}\right]d^{11}.
\end{align}
Similarly, we have
\begin{align}\label{M22}
{\cal M}_{22}
\sim{}&\sum_{\ell=0}^{N-1}\Lambda_{22}^{N-1-\ell}{\cal A}_0^\ell
b_0 +\sum_{\ell=0}^{N-1}\sum_{k=\ell+1}^{N-1}
\sum_{k_1=0}^{k-\ell}\cdots\sum_{k_\ell=0}^{k-\ell-k_1-\cdots-k_{\ell-1}}
\Lambda_{22}^{N-1-\ell}\left({\Lambda_{11}\over\Lambda_{22}}\right)^{k-\ell}({\cal
C}^{11})^\ell d^{11}
\nonumber\\
={}&(\Lambda_{22}-{\cal A}_0)^{-1}(\Lambda_{22}^N-{\cal A}_0^N)b_0
\nonumber\\&
+\Lambda_{22}^{N-1}
\left\{
{1+{{\cal C}^{11}\over\Lambda_{11}}\over1+{{\cal C}^{11}\over\Lambda_{11}}-{\Lambda_{22}\over\Lambda_{11}}}
\left[\left({\Lambda_{11}\over\Lambda_{22}}\right)^{N-1}\left(1+{{\cal C}^{11}\over\Lambda_{11}}\right)^{N-1}-1\right]
-{{{\cal C}^{11}\over\Lambda_{22}}\over1-{{\cal C}^{11}\over\Lambda_{22}}}\left[1-\left({{\cal C}^{11}\over\Lambda_{22}}\right)^{N-1}\right]
\right\}d^{11},
\end{align}
and
\begin{align}\label{M12}
{\cal M}_{12}
\sim{}&\sum_{\ell=0}^{N-1}\Lambda_{12}^{N-1-\ell}{\cal A}_0^\ell
b_0 +\sum_{\ell=0}^{N-1}\sum_{k=\ell+1}^{N-1}
\sum_{k_1=0}^{k-\ell}\cdots\sum_{k_\ell=0}^{k-\ell-k_1-\cdots-k_{\ell-1}}
\Lambda_{12}^{N-1-\ell}({\cal C}^{12})^\ell d^{12}
\nonumber\\
=\,&(\Lambda_{12}-{\cal
A}_0)^{-1}(\Lambda_{12}^N-{\cal A}_0^N)b_0
\nonumber\\& -\Lambda_{12}^N
({\cal C}^{12})^{-1}\left(1+{{\cal
C}^{12}\over\Lambda_{12}}\right)\left[
1-\left(1+{{\cal
C}^{12}\over\Lambda_{12}}\right)^{N-1}
\right]d^{12}
-\Lambda_{12}^{N-2}{\cal C}^{12}\left(1-{{\cal
C}^{12}\over\Lambda_{22}}\right)^{-1}\left[
1-\left({{\cal
C}^{12}\over\Lambda_{12}}\right)^{N-1}
\right]d^{12}.
\end{align}

For weak dissipation $\gamma \tau {}\ll{}1$, the matrix elements
of ${\cal C}^{ab}$ (and ${\cal A}_0$) are of the order of $\gamma
\tau$ or higher [see (\ref{calC})], while the maximum (in
magnitude) eigenvalue $\lambda_1$ is expected to be of order unity
$\lambda_1\sim O(1)$ with corrections of order $\gamma\tau$. In
this case, the dominant contributions to ${\cal M}_{ij}$ are
estimated to be
\begin{equation}\label{approxMij}
{\cal M}_{11} \sim(N-1)\Lambda_{11}^{N-1}d^{11}, \quad {\cal
M}_{22} \sim\Lambda_{11}^{N-1}d^{11}, \quad {\cal M}_{12}
\sim(N-1)\Lambda_{12}^{N-1}d^{12}.
\end{equation}
Therefore, if the above $\rho_N$ (\ref{rhoN2}) is
suitably normalized, the state is approximated as a pure state
\begin{equation}\label{rhoF}
\rho_N\sim|u_1\rangle\langle u_1|+O(\gamma\tau,
\,|\lambda_2/\lambda_1|^{N}),
\end{equation}
for a large, but not extremely large $N$ under which the above approximation is valid.

Notice that the validity of the above expression (\ref{approxMij}) and therefore that of the ensuing relation (\ref{rhoF}) are limited.
Actually for a larger $N$, $N\gamma \tau\gg1$, the approximation $(1+x)^N\sim1+Nx$ with $x=O(\gamma \tau)$, on which the expression (\ref{approxMij}) has been based, is no longer valid.
Instead, since
\begin{equation}
(1+x)^N=e^{N\log(1+x)}\sim e^{N(x-x^2/2+\cdots)},
\end{equation}
we expect, for a larger $N$, say $1/\gamma \tau<N<1/(\gamma \tau)^2$,
\begin{equation}\label{approxMij2}
{\cal M}_{11}
\sim\Lambda_{11}^{N-1}e^{(N-1){{\cal C}^{11}\over\Lambda_{11}}}
({\cal C}^{11})^{-1}d^{11},
\quad
{\cal M}_{22}
\sim\Lambda_{11}^{N-1}e^{(N-1){{\cal C}^{11}\over\Lambda_{11}}}d^{11},
\quad
{\cal M}_{12}
\sim\Lambda_{12}^{N-1}({\cal C}^{12})^{-1}e^{(N-1){{\cal C}^{12}\over\Lambda_{12}}}d^{12}.
\end{equation}
Even though the coefficient ${\cal M}_{11}$ multiplying the term $|u_1\rangle\langle u_1|$ gives the major contribution, the others coefficients are of the same order as the former and never decrease as we increase $N$.
It is easily understood that a similar expression can be found for an even larger $N$.
That is, we are unable to reach a pure state by increasing the number of measurements (projections) $N$, which is in accord with the theorem shown in \cite{ref:PrevPaper}.

A few comments are in order at this point. The pure state
$|u_1\rangle\langle u_1|$ approximately extracted above in
(\ref{rhoF}) for an  intermediate $N$ is {\it not\/} an eigenstate
of the projected dynamics (\ref{rho1}). Actually it is possible to
prove that no pure state can be an eigenstate of such a positive
dynamical map, provided that it includes more than two (in the
case of two dimensions) projections that (are necessarily not
orthogonal and) project to different pure states. Nevertheless, it
is still possible to consider such a situation where effects of
all but one such projections decrease more quickly than that of
one particular projection and as a result for an appropriately
large $N$ the dynamical map effectively becomes single
dimensional, i.e., an effective manifestation of purification.
This is the case in (\ref{rhoF}). Second, in the
strong-dissipation limit ($\gamma\to\infty$) many simplifications
are expected to occur in the general form, but the essential
points are intact. The density matrix $\rho_N$ takes exactly the
same form as the rhs of (\ref{rhoN2}) supplemented with
(\ref{2ndterm}). It is easily seen that $G(\rho_k)=0,\,\forall k$
and the column vector appearing in (\ref{2ndterm}), ${\cal
C}^{a_1b_1}\cdots{\cal C}^{a_\ell
b_\ell}d^{a_{\ell+1}b_{\ell+1}}$, has only the upper component
\begin{equation}
{\cal C}^{a_1b_1}\cdots{\cal C}^{a_\ell b_\ell}d^{a_{\ell+1}b_{\ell+1}}
=\left[
\begin{matrix}
({\cal C}^{a_1b_1})_{11}\cdots({\cal C}^{a_\ell b_\ell})_{11}(d^{a_{\ell+1}b_{\ell+1}})_1\\
\noalign{\medskip}
0
\end{matrix}
\right].
\end{equation}
The asymptotic form of the density matrix is easily evaluated
\begin{align}
\rho_N\sim{}&
N|\lambda_1|^2|u_1\rangle\langle u_1|\Bigl(\langle v_1|\rho_{\downarrow\downarrow}|v_1\rangle,\,\langle v_1|\rho_{\alpha^*}|v_1\rangle\Bigr)({\cal C}^{11})^{N-2}d^{11}\nonumber\\
&
+N|\lambda_1|^2|u_1\rangle\langle u_2|\Bigl(\langle v_1|\rho_{\downarrow\downarrow}|v_2\rangle,\,\langle v_1|\rho_{\alpha^*}|v_2\rangle\Bigr)({\cal C}^{12})^{N-2}d^{12}+h.c.\nonumber\\
&+(N-1)|\lambda_1|^2|u_2\rangle\langle u_2|\Bigl(\langle v_2|\rho_{\downarrow\downarrow}|v_2\rangle,\,\langle v_2|\rho_{\alpha^*}|v_2\rangle\Bigr)({\cal C}^{11})^{N-2}d^{11}
+(N-1)|\lambda_1|^2\Bigl(\rho_{\downarrow\downarrow},\,\rho_{\alpha^*}\Bigr)({\cal C}^{11})^{N-2}d^{11}\nonumber\\
\sim{}&
N|\lambda_1|^2\biggl[|u_1\rangle\langle u_1|\langle v_1|\rho_{\downarrow\downarrow}|v_1\rangle({\cal C}^{11})_{11}^{N-2}(d^{11})_1
+|u_1\rangle\langle u_2|\langle v_1|\rho_{\downarrow\downarrow}|v_2\rangle({\cal C}^{12})_{11}^{N-2}(d^{12})_1+h.c.\nonumber\\
&\qquad\qquad
+|u_2\rangle\langle u_2|\langle v_2|\rho_{\downarrow\downarrow}|v_2\rangle({\cal C}^{11})_{11}^{N-2}(d^{11})_1
+\rho_{\downarrow\downarrow}({\cal C}^{11})_{11}^{N-2}(d^{11})_1\biggr].\label{rhoN4}
\end{align}
If we neglect differences between $({\cal C}^{11})_{11}$ and $({\cal C}^{12})_{11}$ and $(d^{11})_1$ and $(d^{12})_1$, the above expression tells nothing but that the state relaxes to the down state $\rho_{\downarrow\downarrow}=|{\downarrow}\rangle\langle{\downarrow}|$.
On the other hand, if the matrix elements, $({\cal C}^{11})_{11}$ and $(d^{11})_1$, are greater than the others, $({\cal C}^{12})_{11}$ and $(d^{12})_1$, the formers overwhelm the latters as $N$ increases and the density matrix will relax to a mixed state
\begin{equation}
\rho_N\ \longrightarrow\  |u_1\rangle\langle u_1|\langle v_1|\rho_{\downarrow\downarrow}|v_1\rangle+|u_2\rangle\langle u_2|\langle v_2|\rho_{\downarrow\downarrow}|v_2\rangle+\rho_{\downarrow\downarrow}.
\end{equation}
If the opposite case were possible, we would have been given another matrix
\begin{equation}
\rho_N\ \longrightarrow\ \frac{|u_1\rangle\langle
u_2|+|u_2\rangle\langle u_1|}{\langle u_2|u_1\rangle+\langle
u_1|u_2\rangle}
\end{equation}
but it turns out that this is impossible, for this matrix can be shown to have a negative eigenvalue (the trace of the square of this matrix is shown to exceed 1!) and therefore it will never show up in our physical process.
In all cases with strong dissipation, no nontrivial pure state is shown to be extracted, even at an intermediate stage with a finite number of measurements.
Notice that these observations are actually consistent with the exact result;
in the strong-dissipation limit $\gamma\to\infty$, the operator $C_2$ becomes irrelevant [see (\ref{eq:rho(tau)})] and since the operators $C_0$ and $C_1$ project qubit S to the down state, the simultaneous eigenstate should be the down state ($\alpha=0$), for which the operator $V$ becomes diagonal.
This means that $|u_{1(2)}\rangle\to|{\downarrow}({\uparrow})\rangle$ and $\rho_N\to\rho_{\downarrow\downarrow}$.

Finally, it is to be noticed that the dissipative environment also
affects the yield, that is, the probability of obtaining the target pure
state. Actually, the probability of obtaining the above pure
state~(\ref{rhoF}) is considered to be almost equal to the success
probability, which is given by the normalization factor of
$\rho_N$ in (\ref{rhoN2}), if the latter is dominated by the pure
state as in~(\ref{rhoF}), and this factor is of order $\sim
(1+N\gamma\tau)\Lambda_{11}^N$ for an intermediate $N$, e.g.,
$1\ll N<1/\gamma\tau$. On the contrary, the dephasing environment
causes no effect on the yield.

\section{Summary and discussions}\label{sec_final}
We have shown in this paper that the repeated-measurement-based
purification scheme is robust against a dephasing environment and
the up state of qubit S, $|{\uparrow}\rangle\langle{\uparrow}|$,
can be realized asymptotically without any loss of probability,
when qubit X in (rotating-wave) interaction with the former is
repeatedly confirmed to be in the up state,
$|{\uparrow}\rangle_{\rm X}$, just as in the ideal case. In order
to reach the final pure state with fewer steps, we just adjust
parameters, say, the interval between measurements $\tau$, and
minimize $|\cos E_+\tau+i(\omega-\Omega)(2E_+)^{-1}\sin E_+\tau|$,
\begin{equation}
\cos^2\!E_+\tau+{(\omega-\Omega)^2\over4E_+^2}\sin^2\!E_+\tau
=1-{4g^2\over(\omega-\Omega)^2+4g^2}\sin^2\!E_+\tau\ge{(\omega-\Omega)^2\over(\omega-\Omega)^2+4g^2},
\end{equation}
the equality of which is attained when $\cos E_+\tau=0$, realizing an optimal case.
This condition is the same as that for the ideal case.

The reason why the dephasing does not affect the ability of this
kind of purification scheme may be understood in the following
way. The dephasing environment surely disturbs the phases of both
qubits but causes no transitions between up and down states. On
the other hand, our purification scheme is dependent on the
probability of finding a quantum system in some definite state and
the change in its phase has no relevance to this scheme. This is a
naive interpretation of why the ability of this purification
scheme is not affected by the presence of dephasing environment,
when the state $|{\uparrow}\rangle$ is measured.

We have next considered the case of dissipative environment and
examined the ability of the purification scheme, though it is
already known~\cite{ref:PrevPaper} that no (nontrivial) pure state
can be extracted in this case according to the general theorem.
When the qubit X is repeatedly confirmed to be in a definite
state, the target system S, in (rotating-wave) interaction with X,
is forced to be in a definite state, while the surrounding
environment constantly drives the system to the equilibrium. It
seems that a kind of competition between two tendencies, one
forced by the projection and the other relaxing to an equilibrium
state, results in an approximate extraction of a dominant pure
state at an intermediate stage with a large but not extremely
large number of measurements. For a weak damping case, we are able
to find an asymptotic expression of the state of qubit S, which
shows that the dominant contribution is given by one of the
eigenstates of the operator $V$ (belonging to the largest
eigenvalue in magnitude), which is a kind of projected evolution
operator incorporating partly dissipative dynamics and reduces to
the usual unitary operator supplemented with projection when there
is no dissipation. Of course, if we repeat the measurement
indefinitely, the dissipative dynamics would swiftly overwhelm the
effect of projection (purification) and the system would never be
driven to a (nontrivial) pure state, which is in accord with the
general theorem.

It is worth mentioning that in actual situations with various causes of decoherence, one has to be careful about and aware of the presence of such nonideal elements and should not perform measurements indefinitely, for there is a possibility that an optimal result can be attained at a finite number of measurements, as in our simplified model, even though a general criterion seems to be quite difficult to be obtained at present.

\section{Acknowledgments}
This work is partly supported by the bilateral Italian-Japanese
Projects II04C1AF4E on ``Quantum Information, Computation and
Communication'' of the Italian Ministry of Instruction, University
and Research, and 15C1 on ``Quantum Information and Computation''
of the Italian Ministry for Foreign Affairs, by a Grant for The
21st Century COE Program (Physics of Self-Organization Systems) at
Waseda University from the Ministry of Education, Culture, Sports,
Science and Technology, Japan, and by a Grant-in-Aid for
Scientific Research (C) (No.~18540292) from the Japan Society for
the Promotion of Science. Moreover, the authors acknowledge
partial support from University of Palermo in the context of the
bilateral agreement between University of Palermo and Waseda
University, dated May 10, 2004.

\appendix
\section{Elements of ${\cal C}^{ab}$ and $d^{ab}$}\label{app_Cabdab}
Here we show the elements of ${\cal C}^{ab}$ and $d^{ab}$ explicitly
\begin{align}
({\cal C}^{ab})_{11}
&={1-e^{-\gamma \tau}\over2(1+|\alpha|^2)}\langle\alpha^*|u_a\rangle\langle v_a|{\downarrow}\rangle\langle{\downarrow}|v_b\rangle\langle u_b|\alpha^*\rangle+{|\alpha|^2(1-e^{-\gamma \tau}-\gamma \tau e^{-\gamma \tau})\over(1+|\alpha|^2)^2}\langle{\uparrow}|u_a\rangle\langle v_a|{\downarrow}\rangle\langle{\downarrow}|v_b\rangle\langle u_b|{\uparrow}\rangle,\nonumber\\
({\cal C}^{ab})_{12}
&={1-e^{-\gamma \tau}\over2(1+|\alpha|^2)}\langle\alpha^*|u_a\rangle\langle v_a|\alpha^*\rangle\langle\alpha^*|v_b\rangle\langle u_b|\alpha^*\rangle+{|\alpha|^2(1-e^{-\gamma \tau}-\gamma \tau e^{-\gamma \tau})\over(1+|\alpha|^2)^2}\langle{\uparrow}|u_a\rangle\langle v_a|\alpha^*\rangle\langle\alpha^*|v_b\rangle\langle u_b|{\uparrow}\rangle,\nonumber\\
({\cal C}^{ab})_{21}
&={|\alpha|^2\gamma \tau e^{-\gamma \tau}\over2(1+|\alpha|^2)}
\langle{\uparrow}|u_a\rangle\langle v_a|{\downarrow}\rangle\langle{\downarrow}|v_b\rangle\langle u_b|{\uparrow}\rangle,\nonumber\\
({\cal C}^{ab})_{22}
&={|\alpha|^2\gamma \tau e^{-\gamma \tau}\over2(1+|\alpha|^2)}
\langle{\uparrow}|u_a\rangle\langle v_a|\alpha^*\rangle\langle\alpha^*|v_b\rangle\langle u_b|{\uparrow}\rangle,\label{calC}\displaybreak[0]\\
(d^{ab})_1
&={1-e^{-\gamma \tau}\over2(1+|\alpha|^2)}\langle\alpha^*|u_a\rangle\langle v_a|\rho(0)|v_b\rangle\langle u_b|\alpha^*\rangle
+{|\alpha|^2(1-e^{-\gamma \tau}-\gamma \tau e^{-\gamma \tau})\over(1+|\alpha|^2)^2}\langle{\uparrow}|u_a\rangle\langle v_a|\rho(0)|v_b\rangle\langle u_b|{\uparrow}\rangle,
\nonumber\\
(d^{ab})_2
&={|\alpha|^2\gamma \tau e^{-\gamma \tau}\over2(1+|\alpha|^2)}
\langle{\uparrow}|u_a\rangle\langle v_a|\rho(0)|v_b\rangle\langle u_b|{\uparrow}\rangle.
\end{align}

\section{Derivation of the relation (\ref{eq:relation1})}\label{app1}
It is not difficult to perform the first, say, two summations over $k_\ell$ and $k_{\ell-1}$ in the lhs of (\ref{eq:relation1})
\begin{align}
&\sum_{k_1=0}^{k-\ell}\sum_{k_2=0}^{k-\ell-k_1}\cdots
\sum_{k_\ell=0}^{k-\ell-k_1-\cdots-k_{\ell-1}}x_1^{k_1}\cdots x_\ell^{k_\ell}x_{\ell+1}^{k-\ell-k_1-\cdots-k_\ell}\nonumber\\
&\quad
=\sum_{k_1=0}^{k-\ell}\cdots
\sum_{k_{\ell-1}=0}^{k-\ell-k_1-\cdots-k_{\ell-2}}x_1^{k_1}\cdots x_{\ell-1}^{k_{\ell-1}}\left({x_\ell^{k-(\ell-1)-k_1-\cdots-k_{\ell-1}}\over x_\ell-x_{\ell+1}}+{x_{\ell+1}^{k-(\ell-1)-k_1-\cdots-k_{\ell-1}}\over x_{\ell+1}-x_\ell}\right)
\nonumber\\
&\quad
=\sum_{k_1=0}^{k-\ell}\cdots
\sum_{k_{\ell-2}=0}^{k-\ell-k_1-\cdots-k_{\ell-3}}x_1^{k_1}\cdots x_{\ell-2}^{k_{\ell-2}}\left(
{x_{\ell-1}^{k-(\ell-2)-k_1-\cdots-k_{\ell-2}}\over(x_{\ell-1}-x_{\ell+1})(x_{\ell-1}-x_\ell)}+{x_\ell^{k-(\ell-2)-k_1-\cdots-k_{\ell-2}}\over(x_\ell-x_{\ell+1})(x_\ell-x_{\ell-1})}\right.\nonumber\\
&\qquad
\qquad\qquad\qquad\qquad\qquad\qquad\qquad\qquad\qquad\qquad\qquad\qquad\qquad\qquad
\left.+{x_{\ell+1}^{k-(\ell-2)-k_1-\cdots-k_{\ell-2}}\over(x_{\ell+1}-x_\ell)(x_{\ell+1}-x_{\ell-1})}\right).
\end{align}
These expressions are already quite suggestive.
Actually we easily perform the summation of the following form to get
\begin{align}
&\sum_{k_m=0}^{k-\ell-k_1-\cdots-k_{m-1}}x_m^{k_m}
\left({x_{m+1}^{k-m-k_1-\cdots-k_m}\over(x_{m+1}-x_{\ell+1})(x_{m+1}-x_\ell)\cdots(x_{m+1}-x_{m+2})}+\cdots\right.\nonumber\\
&\qquad\qquad\qquad\qquad\qquad\qquad\qquad\qquad\qquad\left.
+{x_{\ell+1}^{k-m-k_1-\cdots-k_m}\over(x_{\ell+1}-x_\ell)(x_{\ell+1}-x_{\ell-1})\cdots(x_{\ell+1}-x_{m+1})}\right)\nonumber\\
&\qquad\qquad
=-x_m^{k-\ell-k_1-\cdots-k_{m-1}+1}
  \left({x_{m+1}^{\ell-m}\over(x_{m+1}-x_{\ell+1})\cdots(x_{m+1}-x_m)}+\cdots
  +{x_{\ell+1}^{\ell-m}\over(x_{\ell+1}-x_\ell)\cdots(x_{\ell+1}-x_m)}\right)\nonumber\\
&\qquad\qquad\qquad\qquad\qquad\qquad\qquad\qquad
+{x_{m+1}^{k-(m-1)-k_1-\cdots-k_{m-1}}\over(x_{m+1}-x_{\ell+1})\cdots(x_{m+1}-x_m)}
+\cdots+{x_{\ell+1}^{k-(m-1)-k_1-\cdots-k_{m-1}}\over(x_{\ell+1}-x_\ell)\cdots(x_{\ell+1}-x_m)}.
\label{eq:1stterm}
\end{align}
The quantity in the square brackets in the first line of the rhs can be written as
\begin{align}
&{x_{m+1}^{\ell-m}\over(x_{m+1}-x_{\ell+1})\cdots(x_{m+1}-x_m)}+\cdots
 +{x_{\ell+1}^{\ell-m}\over(x_{\ell+1}-x_\ell)\cdots(x_{\ell+1}-x_m)}\nonumber\\
&\qquad
={1\over\prod\limits_{\ell+1\ge i>j\ge m}(x_i-x_j)}
\left[x_{\ell+1}^{\ell-m}\prod_{i>j\atop i,j\not=\ell+1}(x_i-x_j)-x_\ell^{\ell-m}\prod_{i>j\atop i,j\not=\ell}(x_i-x_j)+\cdots+(-1)^{\ell-m}x_{m+1}^{\ell-m}\prod_{{i>j\atop i,j\not=m+1}}(x_i-x_j)\right].
\end{align}
Since there can be no singularities at $x_i=x_j$ ($i,j\not=m$) and the quantity in the last square brackets is a polynomial of $(\ell-m)$th order all in $x_{\ell+1},\ldots,x_{m+1}$, one can easily deduce its form as
\begin{align}
f(x_m)(x_{\ell+1}-x_\ell)&(x_{\ell+1}-x_{\ell-1})
\cdots(x_{\ell+1}-x_{m+1})\nonumber\\
&\times(x_\ell-x_{\ell-1})(x_\ell-x_{\ell-2})\cdots(x_\ell-x_{m+1})
\nonumber\\
&\qquad\qquad\qquad\qquad\qquad
\times\cdots(x_{m+2}-x_{m+1}).
\end{align}
The remaining function $f(x_m)$ can be fixed, for example, by the behavior around $x_{\ell+1}=x_m$, to be $f(x_m)=(-1)^{\ell-m}x_m^{\ell-m}$.
This means that the first term in (\ref{eq:1stterm}) is also expressed as
\begin{equation}
-x_m^{k-\ell-k_1-\cdots-k_{m-1}+1}{f(x_m)\over\prod\limits_{\ell+1\ge i>m}(x_i-x_m)}
={x_m^{k-(m-1)-k_1-\cdots-k_{m-1}}\over(x_m-x_{\ell+1})(x_m-x_\ell)\cdots(x_m-x_{m+1})}.
\end{equation}
The validity of (\ref{eq:relation1}) is now evident.

\section{Dominant terms in the summations (\ref{sumoverk})}\label{app_summations}
The fact that the dominant contributions in (\ref{sumoverk}) are
due, in general, to those terms with the same set of indices
$\{ij\}={}\{a_1b_1\}=\{a_2b_2\}=\cdots=\{a_{\ell+1}b_{\ell+1}\}$
is understood by observing that the expression (\ref{sumoverk}) is
nothing but the result of the summations of the terms with
different powers of complex numbers and that no cancellations are
possible among those terms with no relative phase, while strong
cancellation are expected to occur among other terms with
different phases. Indeed, (\ref{sumoverk}) is nothing but the
result of the summations of the form
\begin{align}
\sum_{k=\ell}^{N-1}\sum_{k_1=0}^{k-\ell}\sum_{k_2=0}^{k-\ell-k_1}\cdots&
\sum_{k_\ell=0}^{k-\ell-k_1-\cdots-k_{\ell-1}}x^{N-1-k}x_1^{k_1}\cdots x_\ell^{k_\ell}x_{\ell+1}^{k-\ell-k_1-\cdots-k_\ell}\nonumber\\
&=\sum_{k=\ell}^{N-1}\sum_{k_1=0}^{k-\ell}\sum_{k_2=0}^{k-\ell-k_1}\cdots
\sum_{k_\ell=0}^{k-\ell-k_1-\cdots-k_{\ell-1}}x^{N-1-\ell}(x_1/x_{\ell+1})^{k_1}\cdots(x_\ell/x_{\ell+1})^{k_\ell}(x_{\ell+1}/x)^{k-\ell}\nonumber\\
&=\sum_{k'=0}^{N-1-\ell}\sum_{k_1'=0}^{k'}\sum_{k_2'=0}^{k_1'}\cdots\sum_{k_\ell'=0}^{k_{\ell-1}'}x^{N-1-\ell}(x_1/x)^{k'}(x_2/x_1)^{k_1'}\cdots(x_{\ell+1}/x_\ell)^{k_\ell'},\label{sumoverk'}
\end{align}
the absolute value of which is expected to be maximum when all complex variables $x,x_1,\ldots,x_{\ell+1}$ are in phase.
Since in the present case the absolute values of $\Lambda_{a_nb_n}$s are bounded $|\Lambda_{a_nb_n}|\le|\Lambda_{11}|=|\lambda_1|^2$ (we assume that there is no degeneracy  $|\lambda_1|>|\lambda_2|$ and that the eigenvalues $\lambda_i$ have a nonvanishing relative phase), dominant contributions come from those terms with $x_1=\cdots=x_{\ell+1}=\Lambda_{11}$
when $x=\Lambda_{11}$ or $x=\Lambda_{22}$ and otherwise with $x_1=\cdots=x_{\ell+1}=x$.
We can thus easily extract the dominant contributions.
Notice that when $x=x_1=\cdots=x_{\ell+1}$, the above summations
\begin{equation}\label{sumrearrange}
\sum_{k=\ell}^{N-1}\sum_{k_1=0}^{k-\ell}\cdots\sum_{k_\ell=0}^{k-\ell-k_1-\cdots-k_{\ell-1}}=
\sum_{k=0}^{N-1-\ell}\,\sum_{k_1=0}^{N-1-\ell-k}\cdots\sum_{k_\ell=0}^{N-1-\ell-k-k_1-\cdots-k_{\ell-1}}=
\mathop{\sum\cdots\sum}\limits_{0\le k,k_1,\ldots,k_{\ell+1}\le N-1-\ell\atop k+k_1+\cdots+k_{\ell+1}=N-1-\ell}
\end{equation}
express nothing but the number of distributions of $N-1-\ell$ identical balls into $\ell+2$ boxes allowing empty boxes, that is, $_NC_{\ell+1}$.
Furthermore, $k=\ell$ case is exceptional for in this case the summations over all $a_i$ and $b_j$ in (\ref{FandG0}) are trivially done.


\end{document}